\documentclass[a4paper,11pt]{article}
\pdfoutput=1 

\usepackage{jinstpub} 

\title{\boldmath Performance of the Multigap Resistive Plate Chambers of the Extreme Energy Events Project}

\author[a,k,*]{D.De Gruttola\note{Corresponding author, email: daniele.degruttola@centrofermi.it}}  
\author[a,b]{M.Abbrescia,}
\author[a,c]{C.Avanzini,}  
\author[a,c]{L.Baldini,} 
\author[a,d]{R.Baldini Ferroli,} 
\author[a,c]{G.Batignani,} 
\author[a,q]{M.Battaglieri,}  
\author[a,e]{S.Boi,} 
\author[a,e,o]{E.Bossini,} 
\author[a,f]{F.Carnesecchi,}
\author[a,g]{A.Chiavassa,} 
\author[a,h]{C.Cicalo,} 
\author[a,f]{L.Cifarelli,} 
\author[a]{F.Coccetti,} 
\author[a,i]{E.Coccia,}  
\author[a,j]{A.Corvaglia,}  
\author[a,k]{S.De Pasquale,} 
\author[a,d]{F.L.Fabbri,} 
\author[p]{V.Frolov,} 
\author[a,g]{L.Galante,}
\author[a,g]{P.Galeotti,}
\author[a,f]{M.Garbini,} 
\author[a,q]{G.Gemme,} 
\author[a,g]{I.Gnesi,}
\author[a]{S.Grazzi,}
\author[a,l]{C.Gustavino,} 
\author[a,f,o]{D.Hatzifotiadou,} 
\author[a,r]{P.La Rocca,}   
\author[a,s]{G.Mandaglio,}  
\author[n]{O.Maragoto Rodriguez,}
\author[m]{G.Maron,}  
\author[a,t]{M.N.Mazziotta,}
\author[a,d]{S.Miozzi,} 
\author[a,f]{R.Nania,} 
\author[a,f]{F.Noferini,} 
\author[a,u]{F.Nozzoli,} 
\author[a,f]{F.Palmonari,} 
\author[a,j]{M.Panareo,} 
\author[a,j]{M.P.Panetta,} 
\author[a,e]{R.Paoletti,} 
\author[n]{W.Park,} 
\author[m]{C.Pellegrino,} 
\author[a,q]{L.Perasso,} 
\author[a,c]{F.Pilo,} 
\author[a,g]{G.Piragino,} 
\author[a,d]{S.Pisano,}
\author[a,r]{F.Riggi,}
\author[a]{G.C.Righini,}  
\author[a,k]{C.Ripoli,}  
\author[a,b]{M.Rizzi,}  
\author[a,f]{G.Sartorelli,}
\author[a,f]{E.Scapparone,} 
\author[a,v]{M.Schioppa,} 
\author[a,c]{A.Scribano,} 
\author[a,f]{M.Selvi,}  
\author[a,h]{S.Serci,} 
\author[a,q]{S.Squarcia,}  
\author[a,q]{M.Taiuti,} 
\author[a,c]{G.Terreni,} 
\author[a,w]{A.Trifir\`{o},} 
\author[a,w]{M.Trimarchi,}  
\author[m]{M.C.Vistoli,} 
\author[a,l]{L.Votano,} 
\author[a,f,o]{M.C.S.Williams,}  
\author[a,n,o]{L.Zheng,} 
\author[a,f,o]{A.Zichichi,}  
\author[a,n,o]{R.Zuyeuski}

\affiliation[a]{Museo Storico
della Fisica e Centro Studi e Ricerche "Enrico Fermi", Roma, Italy}
\affiliation[b]{INFN and Dipartimento Interateneo di Fisica, Universit\`{a} di Bari, Bari, Italy} 
\affiliation[c]{INFN and
Dipartimento di Fisica, Universit\`{a} di Pisa, Pisa, Italy} 
\affiliation[d]{INFN, Laboratori Nazionali di Frascati, Frascati (RM), Italy}
\affiliation[e]{INFN Gruppo Collegato di Siena and Dipartimento di Fisica,
Universit\`{a} di Siena, Siena, Italy}
\affiliation[f]{INFN and Dipartimento di Fisica e Astronomia, Universit\`{a} di Bologna, Bologna,
Italy}
\affiliation[g]{INFN and Dipartimento di Fisica, Universit\`{a} di
Torino, Torino, Italy}
\affiliation[h]{INFN and Dipartimento di Fisica,
Universit\`{a} di Cagliari, Cagliari, Italy}
\affiliation[i]{INFN and Dipartimento di Fisica, Universit\`{a} di
Roma Tor Vergata, Roma, Italy}
\affiliation[j]{INFN and Dipartimento di Matematica e Fisica, Universit\`{a} del Salento, Lecce, Italy}
\affiliation[k]{INFN and
Dipartimento di Fisica, Universit\`{a} di Salerno, Salerno, Italy}
\affiliation[l]{INFN, Laboratori
Nazionali del Gran Sasso, Assergi (AQ), Italy}
\affiliation[m]{INFN CNAF, Bologna, Italy}
\affiliation[n]{ICSC World Laboratory, Geneva, Switzerland}
\affiliation[o]{CERN, Geneva, Switzerland}
\affiliation[p]{JINR Joint Institute for Nuclear Research, Dubna, Russia}
\affiliation[q]{INFN and Dipartimento di Fisica, Universit\`{a} di
Genova, Genova, Italy}
\affiliation[r]{INFN and Dipartimento di Fisica e Astronomia,
Universit\`{a} di Catania, Catania, Italy}
\affiliation[s]{INFN Sezione di Catania and Dipartimento di Scienze Chimiche, Biologiche, Farmaceutiche e Ambientali,
Universit\`{a} di Messina, Messina, Italy}
\affiliation[t]{INFN Sezione di Bari, Bari, Italy}
\affiliation[u]{INFN and ASI Science Data Center, Roma, Italy}
\affiliation[v]{INFN and Dipartimento di Fisica,
Universit\`{a} della Calabria, Cosenza, Italy}
\affiliation[w]{INFN Sezione di Catania and Dipartimento di Scienze Matematiche e Informatiche, Scienze Fisiche e Scienze della Terra, Universit\`{a} di Messina, Messina, Italy} 

\emailAdd{EEE Collaboration email: segreteriaEEE@centrofermi.it}
\abstract{The muon telescopes of the Extreme Energy Events (EEE) Project \cite{eee} are made of three Multigap Resistive Plate Chambers (MRPC). The EEE array is composed, so far, of 59 telescopes and is organized in clusters and single telescope stations distributed all over the Italian territory. They are installed in High Schools with the aim to join research and teaching activities, by involving researchers, teachers and students in the construction, maintenance, data taking and data analysis. The unconventional working sites, mainly school buildings with non-controlled environmental parameters and heterogeneous maintenance conditions, are a unique test field for checking the robustness, the low-ageing features and the long-lasting performance of the MRPC technology for particle tracking and timing purposes. The measurements performed with the EEE array require excellent performance in terms of time and spatial resolution, efficiency, tracking capability and stability. The data from two recent coordinated data taking periods, named Run 2 and Run 3, have been used to measure these quantities and the results are described, together with a comparison with expectations and with the results from a beam test performed in 2006 at CERN.
}


\proceeding{The XIV$^{\text{th}}$ Workshop on Resistive Plate Chambers and related detectors\\
Feb. 19-23/2018 \\
Puerto Vallarta, Jalisco State, MEXICO}
\begin{document}
\maketitle
\flushbottom

\section{Introduction}
\label{sec:intro}
A synchronous sparse array of 59 tracking detectors, each composed of 3 Multigap Resistive Plate Chambers (MRPC), has been built to study Cosmic Rays (CR) and CR-related phenomena in the Extreme Energy Events (EEE) Project. EEE is a project by Centro Fermi (Museo Storico della Fisica e Centro Studi e Ricerche "Enrico Fermi") \cite{CFsite}, in collaboration with INFN (Italian National Institute for Nuclear Physics), CERN and MIUR (the Italian Ministry of Education, University and Research).
The network covers more than 10 degrees in latitude and 11 degrees in longitude and it is being upgraded regularly: 6 new stations are being installed within the end of 2018, leading to a 10$\%$ increase in the number of operating telescopes.\\
\indent Time-correlated events at various distances can be measured by the network with the aim of detecting individual Extensive Air Showers (EAS) \cite{coinc1} or performing more challenging measurements as coincidences between two different correlated EAS \cite{coinc2}.
The EEE network can also address the local properties of the CR flux and its space weather-correlated features \cite{forbushDec2,forbushDec3}, CR flux anisotropies in the sub-TeV energy region \cite{anisotropies} and phenomena related with the upward-going particle flux \cite{upward}.\\
\indent The peculiarity of EEE is the involvement of young students of the Italian High Schools and the consequent outreach and educational impact: 6 telescopes are hosted in INFN and CERN laboratories, while the others are installed in the schools, where students and teachers actively participate to construction, data taking activities, taking care of the operation and maintenance of their telescope. Researchers coordinate and supervise activities, providing support during the construction, installation and use of the detectors. Students and teachers are introduced through seminars, lectures and master-classes to the scientific research community, with the opportunity of understanding how a real experiment works, from the infrastructure development to the data acquisition, analysis and publications of scientific results.
This aspect makes the project turn into a challenge. The unconventional working sites, mainly school buildings with non-controlled environmental parameters and the heterogeneous maintenance conditions, are a unique test field for MRPC technology used for particle tracking and timing purposes. More specifically, the robustness, the low-ageing features and the long-lasting performance of this detector are monitored.
The study of the detector performance has been carried out by selecting a data sample from Run 2 and Run 3 (closed in June 2017).


For the sake of completeness, the whole data set collected since fall 2014 reached 70 billions of muon tracks and in 2017 the network has grown up by a factor almost 8 in terms of number of telescopes wrt. 2007. EEE started its observational activity in 2004 with a set of pilot sites in 7 Italian cities and is the largest and long-living MRPC-based telescopes array, with 59 sites and 14 years of data taking.

A detailed description of the detector and the full results of the analysis on the performance can be found in \cite{eee_perf_paper}, while a summary of the performance will be reported in this paper.

\section{MRPC geometry and signal processing}
\label{sec:mrpc}

\begin{figure}[h!]
\centering
\begin{minipage}[c]{.38\textwidth}
\includegraphics[width=1.2\textwidth]{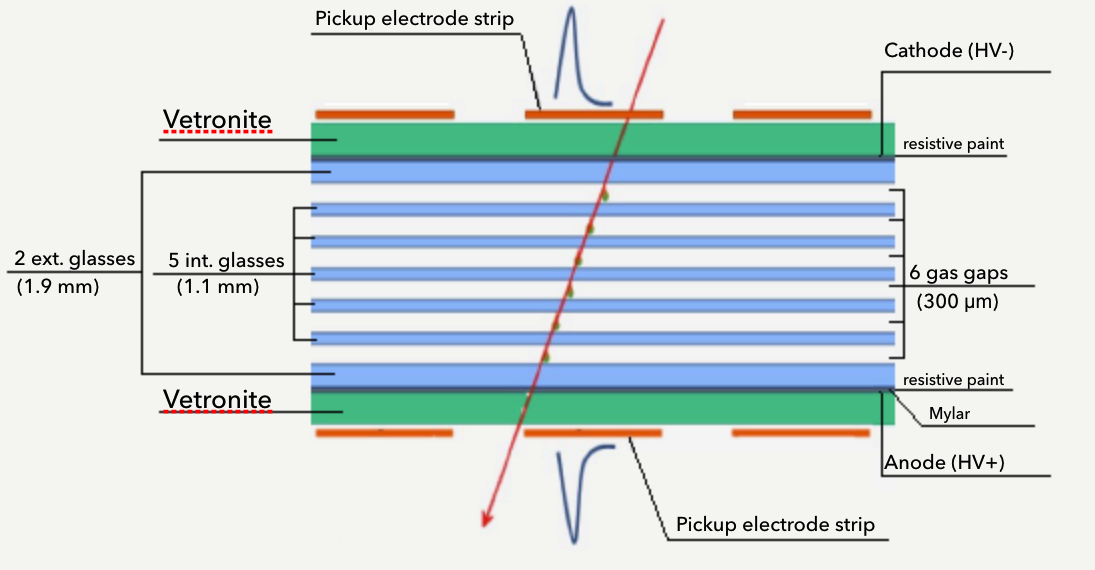}
\end{minipage}%
\hspace{10mm}%
\begin{minipage}[c]{.38\textwidth}
\includegraphics[width=1.2\textwidth]{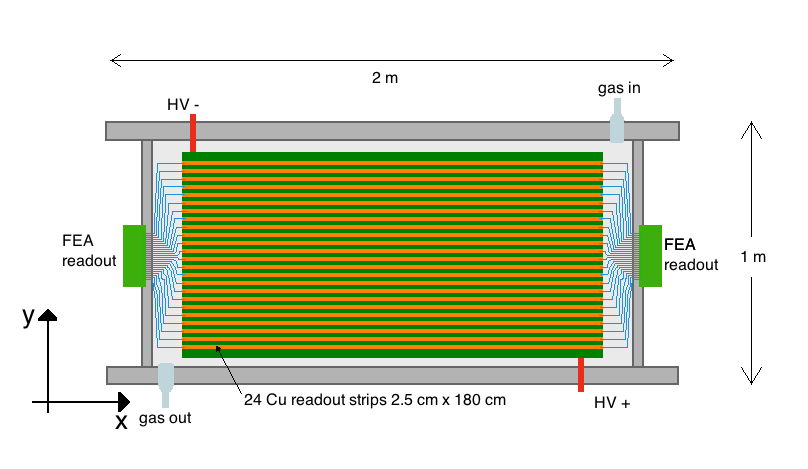}
\end{minipage}
\caption{\footnotesize Left: MRPC inner structure; right: Top view of a MRPC chamber, showing the 24 copper strips, the 2 FEA at the edges of the chamber and the connectors where 2 DC/DC converters providing the HV (top left in blue and bottom right in red) are plugged in. The two figures are not to scale.
}
\label{mrpc}
\end{figure}

The students involvement leads to the need of combining good detector performance, low construction costs and easy assembly procedures.
As a matter of fact, the materials used to build the MRPCs, such as vetronite, honeycomb panel, mylar sheets, glass sheets, fishing line, copper tape and resistive paint are safe, easy to find and to assemble.
High voltage to the chambers, typically in the range 18-20 kV, is provided by a set of DC/DC converters, with output voltage roughly a factor 2000 wrt. the driving low voltage (LV); this solution avoids having usual and possibly dangerous high voltage system provided by HV Power Supply and cables.
The detector structure (Fig. \ref{mrpc} left) consists  of  6  gas  gaps  obtained by stacking glass sheets with voltage applied only to the external surfaces, leaving the inner ones floating. 

The cathode and the anode consist of two glasses (160 cm $\times$ 85 cm, 1.9 mm thick) treated with resistive paint (5-20 M$\Omega$/$\Box$) connected to high voltage, the space between them being divided into the six narrow gaps (300 $\mu$m) by 5 intermediate glass sheets (158 cm $\times$ 82 cm, 1.1 mm thick); inner-glass spacing is assured through a weave made with fishing line. 
Two rigid composite honeycomb panels are used to assure mechanical stability to the whole structure and host, on the two sides, a vetronite panel on whose external surface 24 copper strips (180 cm $\times$ 2.5 cm with a pitch of by 3.2 cm) are laid out to collect the signals induced by particles.
A gas-tight aluminum box, enclosing the whole structure, is filled with a gas mixture consisting of a 98/2 mixture of R134a (C$_{2}$F$_{4}$H$_{2}$) and SF$_{6}$, at a continuous flow of 2 l/h and atmospheric pressure. Promising studies on new gas mixtures are ongoing in order to fulfill with the recent restrictions on greenhouse gases and are reported in the contribution "New Eco-gas mixtures for the Extreme Energy Events MRPCs: results and plans" from S. Pisano.

A schematic top view of a chamber is shown in Fig. \ref{mrpc} right, where in particular the front-end (FEA) boards for the read-out of the strip signals are visible on the short sides. FEA cards incorporate the ultrafast and low power NINO ASIC amplifier/discriminator specifically designed for MRPC operation \cite{asic}.

The 24 copper strips that collect the signal provide two-dimensional information when a cosmic muon crosses the chamber: the $y$ coordinate (short side) is determined by the strip on which the signal is induced, while the $x$ coordinate (long side) is determined by measuring the difference between the arrival time of the signal at the two ends of the strip.
The trigger logic consists in a six-fold coincidence of the OR signals from the FEA cards, whose signals are combined in a VME custom made trigger module. 
The arrival times of the signals are measured using two commercial TDCs (CAEN V1190 - 64 and 128 ch - 100 ps bin). A GPS unit, providing the event time stamp with precision of the order of 40 ns, guarantees the synchronization between telescopes. Data acquisition, monitoring and control are managed by a LabVIEW based program.

A linear fit of the clusters found in the three chambers is performed to reconstruct the tracks and the corresponding $\chi^{2}$ is computed. All possible cluster combinations are used and ordered by their $\chi^{2}$. The track candidates are defined by iteratively selecting the lowest $\chi^{2}$ and removing the corresponding clusters, continuing up to the point when the whole set of available clusters has been assigned to a track. At the end a set of tracks with no hits in common is defined.
The event selection for the measurements presented in this paper requires the rejection of events with more than one track and a track quality given by the cut $\chi^{2}$ < 5.

\section{Time and spatial resolution}
\label{sec:resolution}

\begin{figure}[htb!]
\centerline{
\includegraphics[width=0.5\columnwidth]
{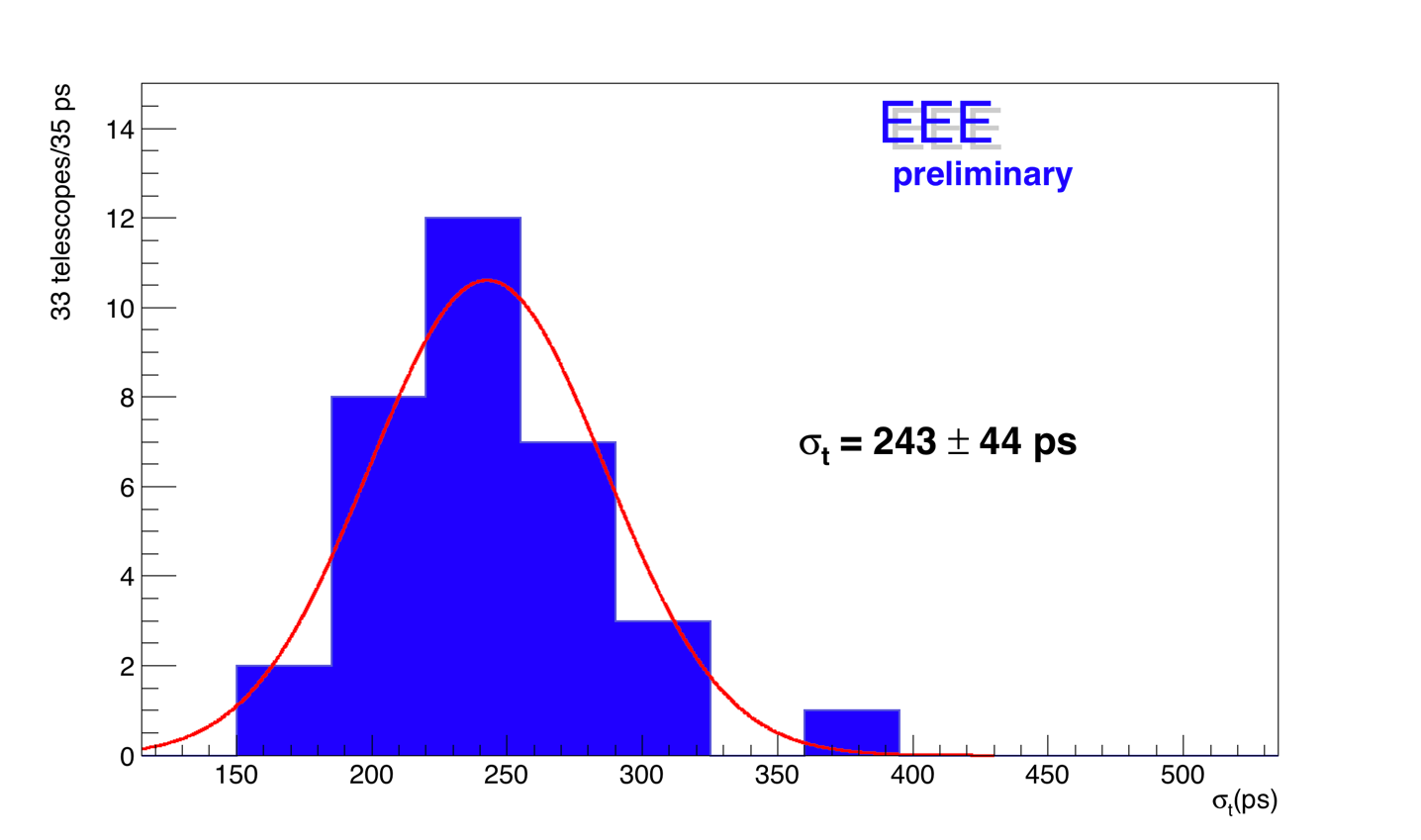}
}
\caption{\footnotesize Time resolution measured with data taken in Run 3, for 33 telescopes; the average time resolution $\sigma_{t}$ = 238 ps is obtained wuth a gaussian fit.}
\label{timeRes_all}
\end{figure}

Data collected during the Run 2 and Run 3 were used to measure time and spatial resolution.
A comparison of the hit information $s$ on the central chamber with the ones on the external chambers is performed, so that the width of the obtained distribution $\Delta s$ is used to estimate the (time/longitudinal spatial/transverse spatial) resolution ($\sigma_{t}$/$\sigma_{x}$/$\sigma_{y}$) of the telescope.
Time (spatial) residuals used for the measurement of the time (spatial) resolution are defined as $\Delta s = (s_{top} + s_{bot})/{2} - s_{mid}$,
where $s_{top}$, $s_{mid}$, $s_{bot}$ are the time (spatial) values for single or clustered hits and $s$ represents $t$, $x$, or $y$ depending on the considered case.
More details can be found in \cite{eee_perf_paper}.\\
A sample of about 8 billion tracks over 31 billions collected during the Run 2 and Run 3 was used to measure $\sigma_{t}$.
To be noted that the time resolution measurement is performed applying Time Slewing correction.
A distribution obtained with the values of time resolution from 33 telescopes of the network is shown in Fig.~\ref{timeRes_all}. A gaussian fit gives an average time resolution $\sigma_{t} = (238 \pm 40) $ ps, that is a value within expectations and totally compatible with EEE specifications. 
The smaller value ($\approx$100 ps) measured at the beam tests performed in 2006 at CERN \cite{beamTest} is explained by the fact that 
in that case conditions are well controlled, with a focused, monochromatic and collinear beam monitored with a set of MultiWire Proportional Chambers (MWPC) and scintillators.

The working points optimization for the next data taking is expected to improve time resolution.
A sample of 2.7 (3.5) billions candidate tracks collected in 30 days from 41 (46)  telescopes in Run 2 (Run 3) has been used to measure $\sigma_{x}$ and $\sigma_{y}$.

\begin{figure}[h!]
\centering
\begin{minipage}[c]{.38\textwidth}
\includegraphics[width=1.2\textwidth]{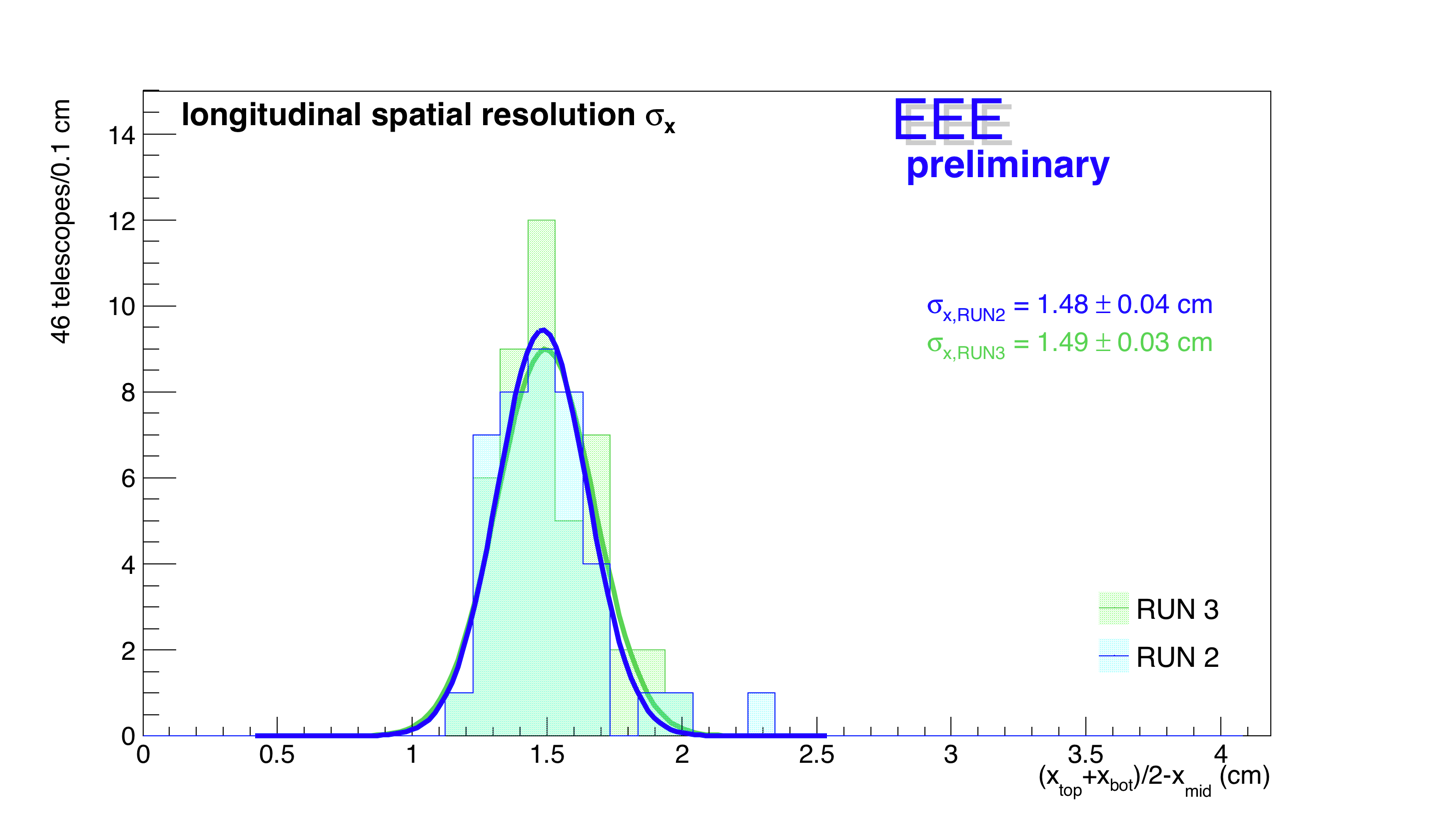}
\end{minipage}%
\hspace{10mm}%
\begin{minipage}[c]{.38\textwidth}
\includegraphics[width=1.2\textwidth]{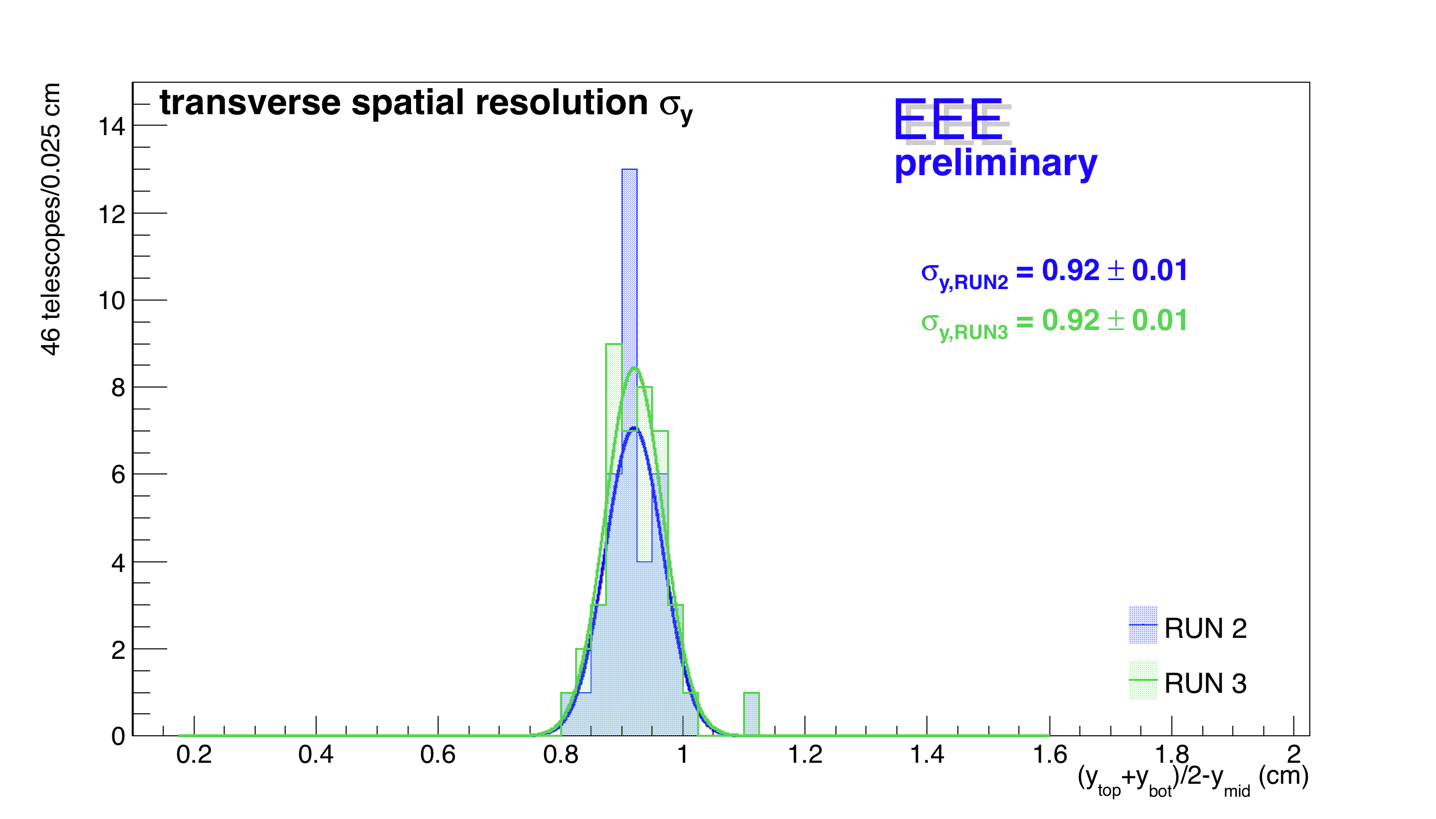}
\end{minipage}
\caption{\footnotesize Longitudinal (left) and transverse (right) spatial resolution measured with data taken in Run 2 and Run 3 with 46 telescopes of the EEE network.
}
\label{res}
\end{figure}

The results from a gaussian fit give an average longitudinal resolution $\sigma_{x_{Run2}} = (1.48 \pm 0.04)$ cm and $\sigma_{x_{Run3}} = (1.49 \pm 0.03)$ cm and an average transverse resolution of $ \sigma_{y_{Run2,Run3}} = (0.92 \pm 0.01)$ cm, in very good agreement with the expectations.
The comparison between the $\sigma_{x}$ (left), or $\sigma_{y}$ (right) distributions from Run 2 and Run 3 are shown in Fig. \ref{res}, where the stability of the network across the two runs is clearly shown.


\section{\textit{Efficiency}}
\label{sec:eff}

\begin{figure}[h!]
\centering
\begin{minipage}[c]{.38\textwidth}
\includegraphics[width=1.2\textwidth]{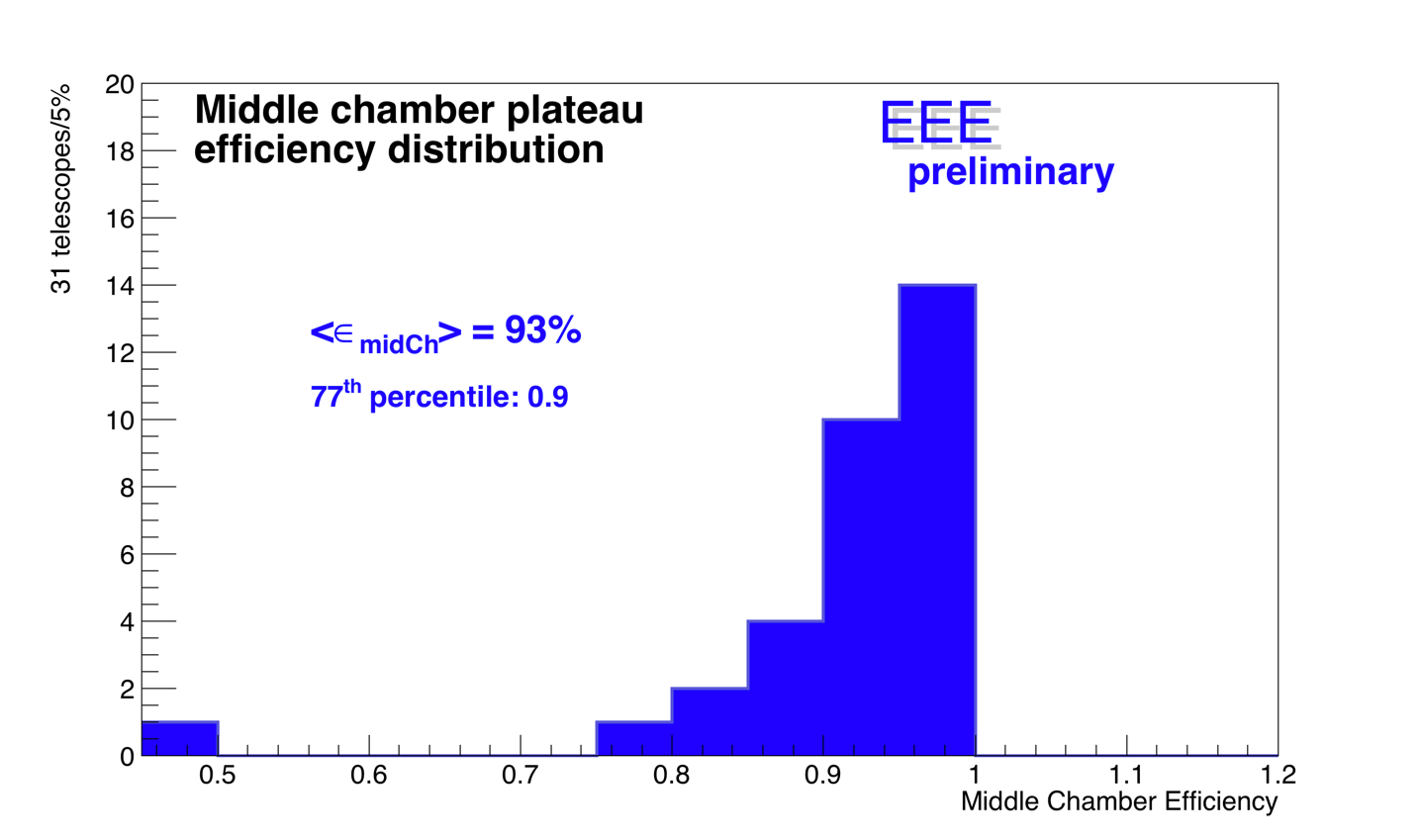}
\end{minipage}%
\hspace{10mm}%
\begin{minipage}[c]{.38\textwidth}
\includegraphics[width=1.2\textwidth]{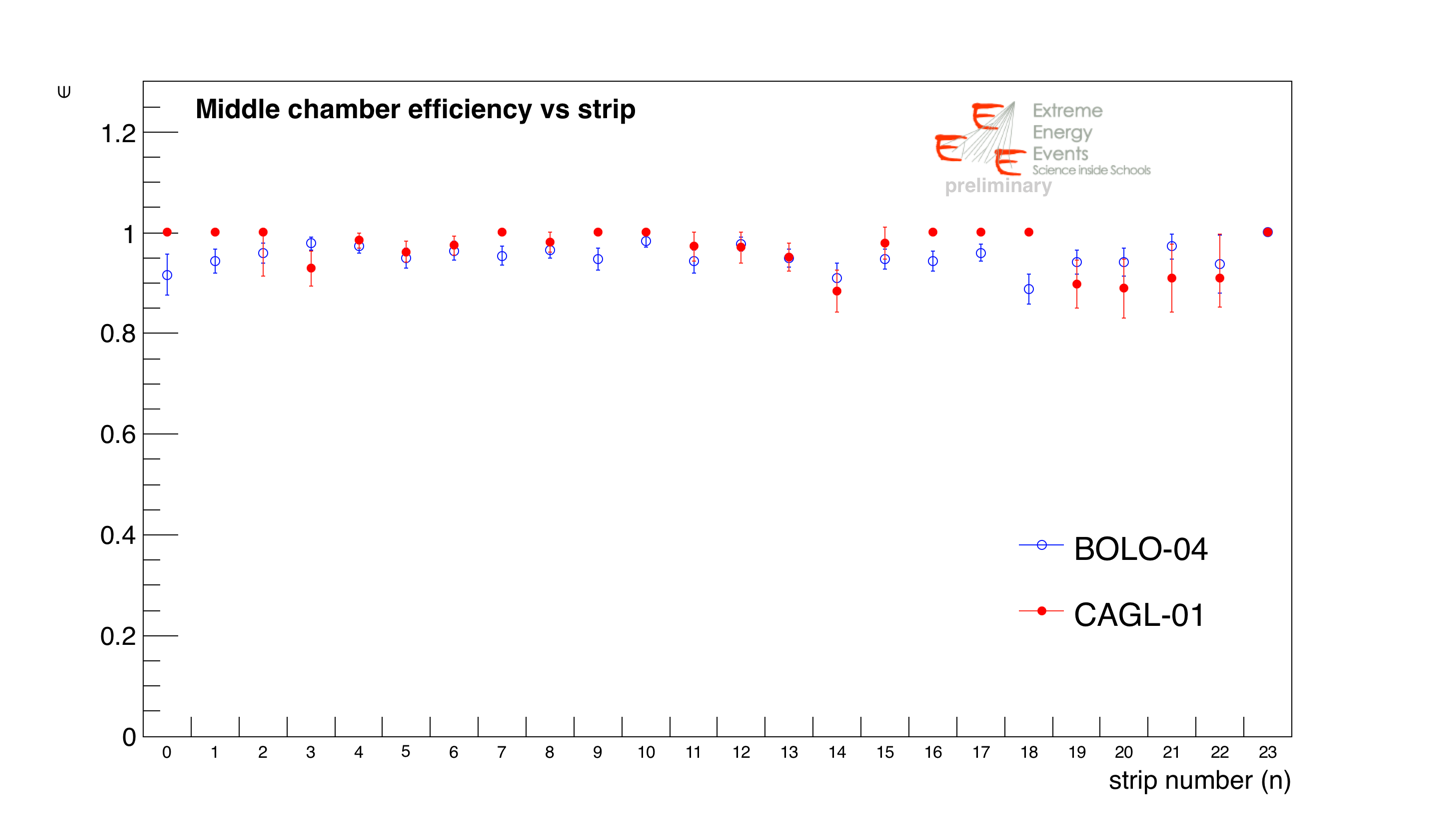}
\end{minipage}
\caption{\footnotesize Left: distribution of the efficiency obtained at the plateau (corrected for standard $p$ and $T$) of the middle MRPC for 31 EEE telescopes. An efficiency better than 90\% is reached by 77\% of the network. Right: efficiency vs. strip for 2 EEE telescopes (located in Emilia Romagna region and Sardinia).
}
\label{eff}
\end{figure}

Efficiency curves as a function of the applied voltage have been measured both at CERN, immediately after chamber construction, and after telescopes installation at schools; in most cases these curves have been obtained using scintillator detectors, employed as external trigger, and with additional electronics. \\
The MRPC efficiency is measured also during the runs by using a slightly modified version of the reconstruction code, with no need of any additional detector.
In particular the trigger logic is changed from the standard 3-chambers operations to a 2-fold coincidence, excluding the chamber under test from the trigger. 
The two chambers in the trigger are also used for tracking and event selection (see \cite{eee_perf_paper} for details). Once a track is defined, the procedure requires to check if a hit is present on the chamber under test within a distance of 7 cm wrt. the expected position.
A HV\footnote{effective high voltage $HV_{eff} = HV \frac{p_{0}}{p} \frac{T}{T_{0}}$, where standard pressure and temperature are set in our case to $p_{0}$ = 1000 mb and $T_{0}$=298.15 $K$.} scan of the chamber is performed, collecting about 150000 events per step.
This method allows to periodically check the detectors performance and provides efficiency values, useful for all analysis. It was applied to the middle chamber, but can be used to measure the efficiency of all the MRPC of a telescope by simply changing the trigger pattern.
A distribution of efficiency values at the plateau from 31 telescopes (middle chamber) obtained from a three parameters sigmoid function fit to each telescope efficiency curve  is shown in Fig. \ref{eff} (left).
The average efficiency of the telescope network is around 93\%, compatible with EEE specs and with results of the beam test in \cite{beamTest}. An efficiency better than 90\% is reached by 77\% of the network, corresponding to 24 telescopes out of 31.
The cause of inefficiency for some telescopes can be related to MRPC ageing or/and dead strips. As an example the efficiency strip by strip for two telescopes of the network is shown in Fig. \ref{eff} (right).

\section{\textit{Long term stability}}
\label{sec:stability}

Data collected by each station are sent to the CNAF center \cite{cnaf}, the computing facility of INFN, where they are stored, reconstructed and made available for analysis.
An automatic on-line Data Quality Monitor (DQM) tool, analyzing a data sample for telescope monitoring purposes, is a crucial tool to achieve long term performance stability, that is not an easy task for a system consisting of detectors hosted in schools, often far away from the nearest technical support. 
Daily report are automatically generated, illustrating the evolution of a set of parameters over the last 48 hours and a fast reaction in case one station deviates from the standard behavior is guaranteed. Using the DQM and full reconstruction outputs, it is also possible to extend such trending plots to longer periods and, as an example, trends for a selection of relevant quantities are reported in \cite{eee_perf_paper} for some telescopes of the EEE network:
In particular these quantities are: the average tracks $\chi^2$, the raw acquisition rate, the average number of hits on the three chambers for each event, the percentage of raw events where at least one track candidate has been found, the average tracks TOF between top and bottom chambers and the rate of events with at least one candidate track.

\section{Conclusions}

The unconventional working sites of the EEE network offer a unique check of the robustness, the ageing  and the long-lasting performance of the MRPC technology. 
These cosmic muon telescopes has been successfully operated in the last 14 years and around 70 billion tracks have been collected during three data taking from 2014 to 2017.
The observatory has grown up by a factor almost 8 in terms of number of telescopes wrt. 2007 and the EEE network is currently the largest and long-living MRPC-based telescopes network, with 59 active sites and 14 years of data taking. 
The results of the analysis on the performance of the network are fully compatible with the EEE requirements in terms of efficiency ($\sim$93\%), time resolution (238 ps) and spatial resolution (1.5 cm and 0.9 cm respectively for longitudinal and transverse direction).
At the moment the EEE Collaboration is focusing on further improvements of the performance in terms of duty cycle and optimization of the working points of the telescopes.



%
%

\end{document}